\documentclass[12pt]{article}

\usepackage{amssymb}
\usepackage{graphicx,psfrag}
\usepackage{enumerate}
\usepackage[superscript]{cite} 
\usepackage{url} 
\usepackage{hyperref}

\usepackage{booktabs}
\newcommand{\blind}{1}
\newcommand{\indep}{\perp \!\!\! \perp}

\usepackage{amsthm}
\usepackage{amsmath}

\newtheorem{proposition}{Proposition}

\begin{document}

\def\spacingset#1{\renewcommand{\baselinestretch}%
{#1}\small\normalsize} \spacingset{1}


\if1\blind
{
  \title{\bf Evaluating Treatment Benefit Predictors using Observational Data: Contending with Identification and Confounding Bias}
  \author{Yuan Xia, Mohsen Sadatsafavi, and Paul Gustafson}\date{}
  \maketitle
} \fi

\if0\blind
{
  \bigskip
  \bigskip
  \bigskip
  \begin{center}
    {\LARGE\bf Evaluating Treatment Benefit Predictors using Observational Data: Contending with Identification and Confounding Bias}
\end{center}
  \medskip
} \fi

\bigskip

\begin{abstract}
A treatment benefit predictor (TBP) is a function that maps patient characteristics to an estimate of the treatment benefit for that patient. Such predictors support optimizing individualized treatment decisions, which are central to precision medicine. However, evaluating the predictive performance of a TBP is challenging, as this often must be conducted in a sample where treatment assignment is not random. After briefly reviewing several metrics for evaluating TBPs, we show conceptually how to evaluate a pre-specified TBP using observational data from the target population, for a binary treatment decision at a single time point. We exemplify with a particular measure of discrimination (the concentration of benefit index) and a particular measure of calibration (the moderate calibration curve). The population-level definitions of these metrics involve the latent treatment benefit variable, but we show identification by re-expressing the respective estimands in terms of the distribution of observable data only. We also show that in the absence of full confounding control, bias propagates in a more complex manner than when targeting more commonly encountered estimands. We find the patterns of biases are often unpredictable, and general intuition about the direction of bias in causal effect estimates does not hold in the present context.

\end{abstract}

\noindent%
\vfill

\newpage
\spacingset{1.45} 

\section*{Introduction}

Precision medicine aims to optimize medical care by tailoring treatment decisions to the unique characteristics of each patient. 
This objective naturally falls in the intersection between predictive analytics and causal inference; the former aims at predicting the outcome of interest given salient patient characteristics, and the latter seeks to answer counterfactual ``what if" questions about the outcome.
Most progress in clinical prediction models has centered around estimating the risk of adverse outcomes, often guiding treatment decisions by prioritizing high-risk individuals for intervention \cite{sperrin2021invited}. However, the success of such guidance relies on an implicit assumption: that those classified as high risk are also those who will benefit the most from the treatment. A more relevant approach for decision-making is to directly predict treatment benefits based on patient characteristics. In this context, treatment benefit refers to the conditional risk reduction for treated individuals given their characteristics, compared to the conditional risk they would face under identical conditions without the treatment. This comparison, which is hypothetical, requires causal reasoning to predict \cite{van2020prediction}. Such a prediction is often termed ``causal prediction'' or ``counterfactual prediction'' \cite{prosperi2020causal}. In causal inference, the treatment benefit we are predicting is the conditional average treatment effect.

We are interested in a function that maps a patient's characteristics to their putative treatment benefit, which we call a treatment benefit predictor (TBP).
Suppose we have a pre-specified TBP developed from a randomized controlled trial (RCT) in one setting, where it demonstrates strong predictive performance. Before being adopted into patient care for the target population with a potentially distinct setting, this TBP needs to be evaluated (validated) in a random sample from the target population \cite{riley2024evaluation}. 
Transportability and generalizability have been studied across epidemiology and various adjacent disciplines, focusing on concepts, assessment methods, and correction techniques \cite{degtiar2023review, dahabreh2020extending, westreich2017transportability, pearl2011transportability}. 
Unlike recent work that aims to correct or improve transportability or generalizability, we do not attempt to adapt or re-train the TBP for the target population. Instead, our goal is to assess how well a TBP developed in a source population performs when applied to a target population as-is. This evaluation may offer practical insight into the inherent transportability or generalizability of the TBP.

Traditionally, performance metrics for risk prediction are categorized into measures of overall fit, discrimination, and calibration \cite{Riley2019, Steyerberg2019}.
Various performance measures for TBPs have been formulated by extending concepts from risk prediction to the treatment-benefit paradigm.
The overall fit of a TBP is typically evaluated by the discrepancy between predicted and estimated treatment benefits \cite{hill2011bayesian, schuler2018comparison}. 
Discriminative ability reflects how well a TBP ranks patients by treatment benefit. The c-for-benefit statistic by van Klaveren et al. extends the c-statistic \cite{VanKlaveren2018}, but its limitations have been noted \cite{efthimiou_measuring_2023, van2023measuring, Hoogland_2024}, including that it does not satisfy the properties of a proper scoring rule \cite{xia_methodological_2023}. Other measures include the concentration of benefit ($C_b$) index \cite{Sadatsafavi2020} and rank-weighted average treatment effect metrics \cite{yadlowsky2024evaluating}, both related to the relative concentration curve \cite{yitzhaki1991concentration} and Qini curve \cite{Radcliffe2007UsingCG}. These metrics assess whether patients prioritized by TBP gain more from treatment than those selected randomly.
Calibration evaluates the agreement between the predicted and the actual quantities. 
It is typically assessed using calibration plots, where predicted benefit is on the x-axis and observed benefit, either smoothed or grouped, is on the y-axis \cite{van_calster_calibration_2016}. For TBPs, calibration has been used to evaluate how well predicted and actual benefits align \cite{xu2022calibration, efthimiou_measuring_2023, Hoogland_2024}.
None of these extensions are straightforward, given the unavailability of the counterfactual outcome. 
Another challenge arises when evaluating TBP performance using observational studies, as these introduce potential confounding bias. 
This is especially relevant when RCTs are unavailable, underpowered, or not representative of the target population.

The primary contributions of this paper are twofold: (1) We show conceptually how to evaluate a TBP from observational data, by building a bridge between causal quantities and observable quantities. 
(2) We show the unpredictable pattern of bias of performance metrics incurred if we are not fully controlling for confounding.
More specifically, for the first contribution, we re-express the metrics, initially defined in terms of the latent treatment benefit variable, in terms of the distribution of observable quantities. Our emphasis is on the conceptual framework, deliberately setting aside estimation details. The resulting foundation will allow for various estimation methods. For the second contribution, we extend the understanding of identification and confounding bias from the context of well-understood estimands such as the average treatment effect to estimands which describe the predictive performance of a pre-specified TBP \cite{imbens2003sensitivity, veitch2020sense}. To illustrate, we focus on two existing performance metrics for assessing pre-specified TBPs: the $C_b$ index and the moderate calibration curve.

\section*{Notation and Assumptions}
Each individual in the target population is described by $(Y^{(0)}, Y^{(1)}, A, X, Z)$ with joint distribution $\mathbb{P}$.
Here $A$ denotes a binary treatment, and $Y^{(a)}$ is the potential outcome that would be observed under treatment $A = a$, where $a \in \{0, 1\}$.
The vector $X$ consists of pre-treatment covariates that will be used to predict treatment benefit in routine clinical practice, i.e., potential effect modifiers. For evaluation, $X$ must also be observable in the observational studies, which may provide a mix of predictive variables and confounders. The vector $Z$ contains additional covariates that, together with $X$, are required to control for confounding, but $Z$ need not be effect modifiers.

Without loss of generality, a larger $Y^{(a)}$ is assumed to be the preferred outcome. The individual treatment benefit of treatment $A = 1$ relative to $A = 0$ is quantified as $B := Y^{(1)} - Y^{(0)}$, which is unobservable.
In clinical practice, the ideal quantity to guide treatment decisions for an individual with $X=x$ is $\tau_0(x) = \operatorname{E}[B \mid X = x]$, which conditions only on the routinely accessible covariates $X$.
We denote the mean treatment benefit for smaller subgroups partitioned by both $X$ and $Z$ as $\tau(x, z) = \operatorname{E}[B \mid X = x, Z = z]$. 
A TBP denoted as $h(x)$, which predicts $\tau_0(x)$,
can guide treatment decision-making. For example, the care provider may offer treatment only to those with $h(x) > 0$.
We denote the predicted treatment benefit from $h(x)$ as $H:= h(X)$, and the best possible TBP is $\tau_0(x)$ itself. 

\begin{figure}[h]
    \centering
    \includegraphics[width=0.35\linewidth]{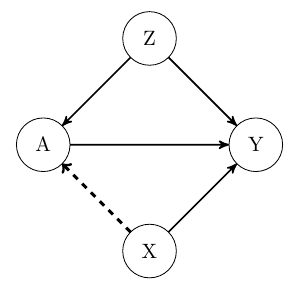}
    \caption{The directed acyclic graph (DAG) shows the causal effect of the exposure ($A$) on the outcome ($Y$). Covariates $X$ are used to build the TBP, while $Z$ is not but is included to control for confounding. The dashed arrow indicates that $X$ may include both confounders and predictors.}
    \label{fig:0}
\end{figure}

Observed data from the observational study are realizations of a vector of random variables drawn from an underlying probability distribution $\mathbb{P}_{obs}$, denoted as $(Y, A, X, Z) \sim \mathbb{P}_{obs}$. Note that $\mathbb{P}_{obs}$ is a consequence of $\mathbb{P}$, as the observed outcome has the form $Y = Y^{(1)}A + Y^{(0)}(1-A)$.
We define $\mu_{a}(x,z) = \operatorname{E}[Y \mid A =a, X = x, Z = z]$ and $\mu_{a}(x) = \operatorname{E}[Y \mid A =a, X = x]$. We denote the propensity score as $e(x, z) = \operatorname{P}(A = 1 \mid X = x, Z = z)$. 

\textbf{Example: Lung function measure.} Consider evaluating pre-specified TBPs that predict the benefit of bronchodilator therapy for lung function measured by forced expiratory volume in 1 second ($\text{FEV}_1$) using observational data. The TBPs considered are functions of obesity and symptom severity ($X = (X_1, X_2)$). These variables are often available in routine primary care and can be used to make treatment decisions.
However, socioeconomic status ($Z$), a confounding variable that needs to be controlled along with obesity and symptom severity, is not often used to predict bronchodilator benefit.
The underlying causal structure is shown in Figure~\ref{fig:0}.

To ensure that $\tau_0(x)$ can be identified from the observed data, $\mu_{a}(x,z)$ must have a causal interpretation. Therefore, the following assumptions are commonly required, though they may not all be strictly necessary in all contexts \cite{greenland2017and}:
(1) no interference: between any two individuals, the treatment taken by one does not affect the potential outcomes of the other;
(2) consistency: the potential outcome under the observed treatment assignment equals the observed outcome $Y$; (3) overlap (positivity): the conditional probability of receiving the active treatment is bounded away from $0$ and $1$, i.e., $0 < e(x,z) < 1$, for all possible $x$ and $z$; and (4) conditional exchangeability: the treatment assignment is independent of the potential outcomes given the variables in $(X, Z)$. Therefore, we assume that the variables in $(X, Z)$ constitute a sufficient adjustment set for identifying the treatment benefit.

\section*{Predictive Performance Metrics}
Among possible metrics for evaluating a pre-specified TBP $h(x)$, we illustrate two by specifying their estimands to clarify how predictive performance can be assessed using observational data.

\subsection*{Discrimination}
The $C_b$ index is a single-value summary of the difference in average treatment benefit between two treatment assignment rules applied to a randomly chosen pair of individuals: `treat at random' and `treat greater $H$.' \cite{Sadatsafavi2020} With independent copies $(B_1, H_1)$ and $(B_2, H_2)$ randomly drawn under $\mathbb{P}$, the $C_b$ index of $H$ is defined as:
\begin{equation}
  C_b = 1 - \frac{\operatorname{E}[B_1]}{\operatorname{E}[B_1I(H_1 \geq H_2) + B_2I(H_1 < H_2)]}, \label{cb}
\end{equation}
where $I(\cdot)$ is the indicator function.
The numerator in Equation (\ref{cb}) operationalizes the strategy of  `treat at random', and the denominator refers to the strategy of `treat greater $H$.'
If two patients have the same $H$, treatment is assigned at random. When $\operatorname{E}[B] > 0$ and $h(x)$ is at least no worse than random, then $0 \leq C_b \leq 1$. The $C_b$ index reflects the percentage reduction in expected benefit when treating at random versus treating the patient with greater $H$; hence higher $C_b$ indicates better discrimination by $h(x)$.

We note that $C_b$ can be expressed in terms of the cumulative distribution function (CDF) of $H$, thereby eliminating the necessity of considering patient pairs to ascertain the expectation in Equation (\ref{cb}). 
For continuous $H$,  we express the $C_b$ as
\begin{align*}
 C_b = 1 - \frac{\operatorname{E}[B]}{2\operatorname{E}[BF_H(H)]},
\end{align*}
where $F_H(\cdot)$ denotes the CDF of $H$. This expression clearly indicates that $C_b$ summarizes the predictive performance of $h(\cdot)$ based on its ranking ability. 
Although $H$ is typically continuous, we derive a general form for discrete $H$ to facilitate simple (illustrative) examples. This form requires a correction term. Appendix S1 derives the corrected form.

\subsection*{Calibration}
The concept of moderate calibration for risk prediction is that the expected value of the outcome among individuals with the same predicted risk is equal to the predicted risk \cite{van_calster_calibration_2016}. 
Similarly, in treatment benefit prediction, a TBP $h(x)$ can be considered moderately calibrated if, for all $h$ in the support of $H$, $$\operatorname{E}[B \mid H = h] = h,$$
where $\operatorname{E}[B \mid H = h]$ is the moderate calibration function.
For example, if $h(x)$ is moderately calibrated and predicts a group of individuals to have $H = 0.5$, we should expect that the average treatment benefit within the group is also $0.5$. 
Furthermore, $h(x)$ is strongly calibrated if $\tau_0(X) = H$.
Calibration of TBPs can also be visualized using calibration plots, which display $\operatorname{E}[B \mid H = h]$ against $h$ \cite{van2019calibration}.
A moderately calibrated TBP yields points on the diagonal identity line.

\subsection*{Re-express Performance Metrics}
These predictive performance metrics for TBP involve the latent variable $B$. To evaluate a pre-specified $h(x)$ using observational data, we need to re-express the metrics through the bridge between causal quantities and observed quantities, given by $\tau(X, Z) = \mu_1(X,Z) - \mu_0(X,Z)$.
For continuous $H$, we re-express $C_b$ as
\begin{align*}
 C_b = 1 - \frac{\operatorname{E}[\mu_1(X,Z) - \mu_0(X,Z)]}{2\operatorname{E}[\left(\mu_1(X,Z) - \mu_0(X,Z)\right)F_H(H)]}.
\end{align*}
Similarly, to evaluate the moderate calibration of any $h(x)$, we have
\begin{align*}
 \operatorname{E}[B \mid H = h] = \operatorname{E}[\mu_1(X,Z) - \mu_0(X,Z)\mid H = h].
\end{align*}

Note that $\tau(x, z)$, which plays a vital role in determining both $C_b$ and $\operatorname{E}[B \mid H = h]$, is identifiable by fully adjusting for confounding.
While the expression $\tau(X, Z) = \mu_1(X,Z) - \mu_0(X,Z)$ aligns well with outcome-based estimation, the framework is broadly applicable and not confined to outcome modeling alone. Alternatively, it can be expressed as 
$\tau(X,Z) = \operatorname{E}\left[\frac{Y(A - e(X, Z))}{e(X, Z)(1 - e(X, Z))} \middle| X, Z\right]$. We choose the outcome-based expression for convenience, recognizing that both expressions are equivalent at the population level. 
However, variations may emerge when considering specific finite-sample estimating techniques associated with each expression. We defer the finite-sample estimation of $C_b$ and of the calibration curve to the Discussion section.

\section*{Confounding Bias of Performance Metrics}

Since $X$ is pre-specified for the TBP, we have defined $Z$ as {\em additional} covariates needed with $X$ to identify the causal quantities, based on the causal structure. However, some confounding may remain if observational data are treated as randomized or if confounders are unmeasured.
Thus, it is crucial to investigate this potential confounding bias and understand how the lack of complete control for confounding factors might affect the accuracy of our TBP performance evaluations.

Specifically, we focus on the special case where we do not control for $Z$, and $X$ alone is insufficient to control for confounding. We denote by $\text{bias}(X)$ the confounding bias as a function of $X$:
\begin{align*}
 \text{bias}(X) = \mu_1(X) - \mu_0(X) - \tau_0(X).
\end{align*}
To illustrate the propagation of \(\text{bias}(X)\) to performance measures, we denote the inaccurate $C_b$ and $\operatorname{E}[B \mid H]$, calculated only controlling for $X$ but not $Z$, as $\tilde{C}_b$ and $\tilde{E}[B \mid H]$, respectively.

The bias of $C_b$ is $\tilde{C}_b - C_b$, and the bias function of the moderate calibration curve is $\tilde{E}[B \mid H = h] - \operatorname{E}[B \mid H = h]$ for all $h$, both influenced by $\text{bias}(X)$.
For $C_b$, non-zero $\text{bias}(X)$ prevents the correct identification of both $\operatorname{E}[B]$ and $\operatorname{E}[BF_H(H)]$, but in different ways. The overall bias of $C_b$ is further complex as it involves the difference between two ratios. For example, it is possible to achieve zero deviation in $C_b$ even with non-zero $\text{bias}(x)$ for all $x$. These factors make determining the direction of the bias of $C_b$ more challenging.

For $\operatorname{E}[B \mid H = h]$, the bias function is the conditional mean of $\text{bias}(X)$ given $H = h$. Therefore, it depends on both $\text{bias}(X)$ and the association between $H$ and $X$. For instance, even if $\text{bias}(x) \neq 0$ for all $x$, the moderate calibration curve can still exhibit zero deviation because $\operatorname{E}[\text{bias}(X) \mid H = h] = 0$ may hold. The behavior of $\operatorname{E}[\text{bias}(X) \mid H = h]$ may be more predictable when it fully aligns with the behavior of $\text{bias}(X)$. 
In the next section, we further investigate these biases in illustrative examples to demonstrate how $\text{bias}(X)$ propagates to the evaluation measures.

\section*{Evaluating TBP in Synthetic Populations}
\subsection*{Synthetic Target Populations}
Building on the lung function measure example, we now illustrate how to evaluate pre-specified TBPs and explore the impact of confounding bias using two synthetic target populations. In the first population, obesity and symptom severity are binary baseline covariates, and socioeconomic status is also binary. There is dependence among the three covariates. The exposure is a binary indicator for bronchodilator therapy, and the outcome is a binary indicator for improvement in $\text{FEV}_1$. 
The individual treatment benefit $B$ of having bronchodilator therapy can take values of $-1, 0$, or $1$, where $-1$ indicates harm, $0$ no benefit, and $1$ benefit. 
The probability of receiving therapy is a function of socioeconomic status, with higher socioeconomic status associated with a higher probability of receiving therapy. The probability of having $\text{FEV}_1$ improvement is affected by all three covariates, but their effects differ between the treatment and control groups. Consequently, the conditional expected treatment effect is a function of the three covariates.
The second population differs slightly from the first one in that $X$ contains only one binary baseline covariate, say symptom severity, and inverse logit functions of covariates are used to describe the probability of receiving therapy and the probability of having $\text{FEV}_1$ improvement.

Both populations are simple enough to allow closed-form expressions for $C_b$ and $\operatorname{E}[B \mid H = h]$, and to trace how confounding bias propagates to bias in $\tilde{C}_b$ and $\tilde{E}[B \mid H]$ for pre-specified TBPs. (See Appendix S2 for the full mathematical setup of the two populations.) Next, we compare evaluation results, and illustrate how $\text{bias}(X)$ and the bias of metrics are influenced by the strength of confounding.
\subsection*{Results}
We formulate four pre-specified TBPs to be evaluated in the first population, each of the form
\begin{align*}
    h(x_1, x_2) &= c_0 + c_1 x_1 + c_2 x_2 + c_3 x_1x_2,
\end{align*}
where $c_0, c_1, c_2$ and $c_3$ are coefficients.
In particular, $h_1(x_1, x_2)$ is the mean of covariates with $c_0 = c_3 = 0$ and $c_1 = c_2 =  0.5$, and $h_2(x_1, x_2)$ is designed to be moderately calibrated by carefully choosing the coefficients.
Let $h_3(x_1, x_2) := \tau_0(x_1, x_2)$, which is strongly calibrated. Let $h_4(x_1, x_2) := \mu_1(x_1, x_2) - \mu_0(x_1, x_2)$, a TBP derived from the target population without adjusting for $Z$.
(See Appendix S3 for the detailed definitions of the coefficients for TBPs.)

\begin{table}[ht!]
  \centering
  \caption{Metric values with and without full confounding control}
  \label{tab:cb}
  \begin{tabular}{l c c c c}
    \toprule
    Metric & $ h_1(x_1,x_2) $ & $ h_2(x_1,x_2) $ & $ h_3(x_1,x_2) $ & $ h_4(x_1,x_2) $ \\
    \midrule
        $C_b$        & 0.6069 & 0.6069 & 0.6298 & 0.6298   \\
    $\tilde{C}_b$     & 0.3307 & 0.3307 & 0.3446 & 0.3446  \\
    \bottomrule
  \end{tabular}
\end{table}

\begin{figure}[ht!]
  \centering
  \includegraphics[scale=0.5]{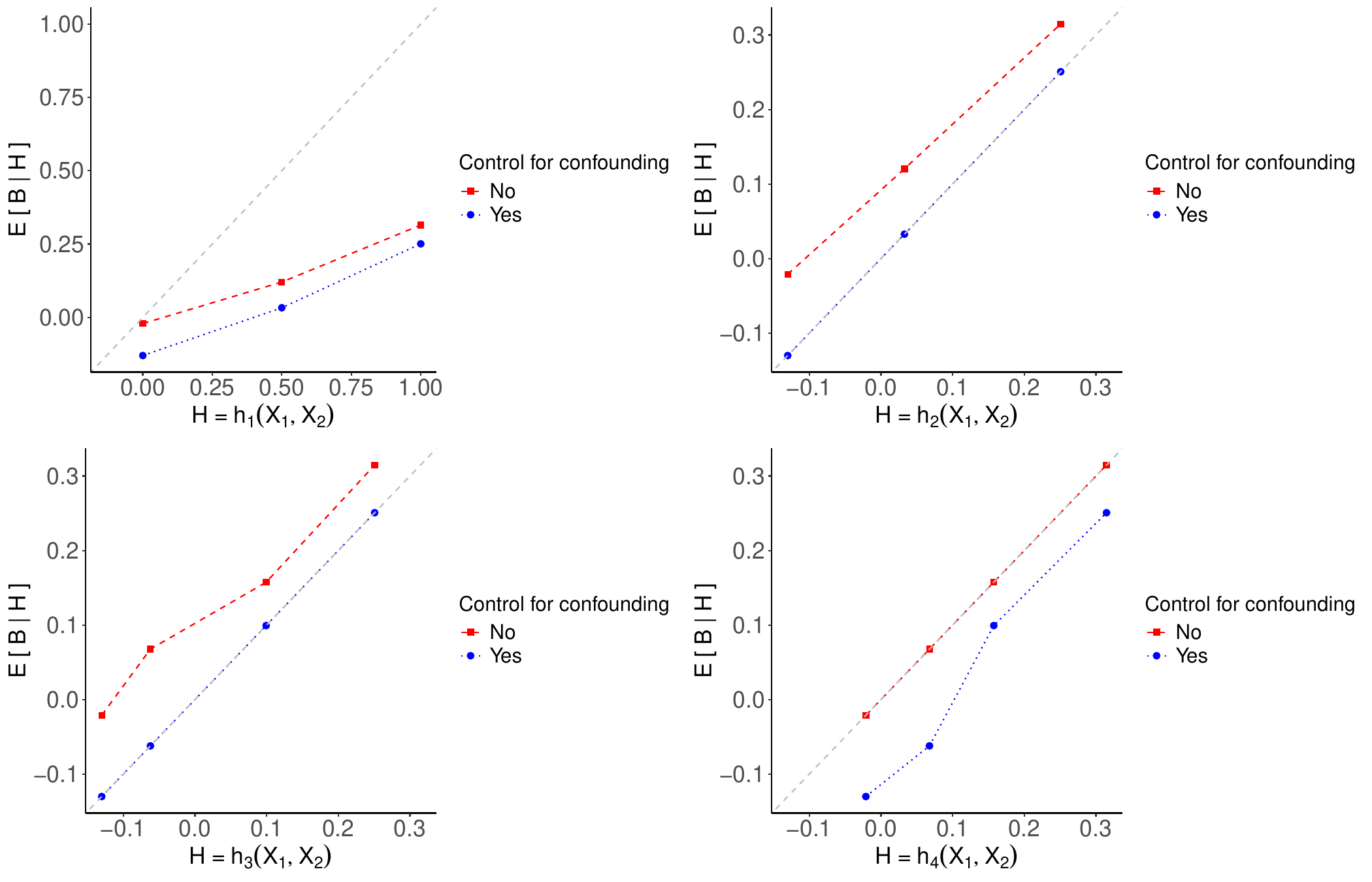}
  \caption[small]{The moderate calibration plots for the four TBPs.
  The blue dotted curves refer to $\operatorname{E}[B \mid H]$, and the red dashed curves refer to $\tilde{\operatorname{E}}[B \mid H]$.}
  \label{fig:1}
\end{figure}

Table~\ref{tab:cb} shows that $h_1$ and $h_2$ have identical discriminatory ability, as do $h_3$ and $h_4$, because the CDFs of $H_1$ and $H_2$ are the same, as are those of $H_3$ and $H_4$. As a result, the corresponding TBPs rank patients identically. Additionally, $h_3$ and $h_4$ yield slightly larger $C_b$ values than $h_1$ and $h_2$, indicating that $H_3$ and $H_4$ produce more effective treatment assignment rules and greater average treatment benefits. Although the values of $\tilde{C}_b$ support the superiority of $h_3$ and $h_4$, they underestimate $C_b$ for all TBPs.

\begin{figure}[ht!]
  \centering
  \includegraphics[scale=0.8]{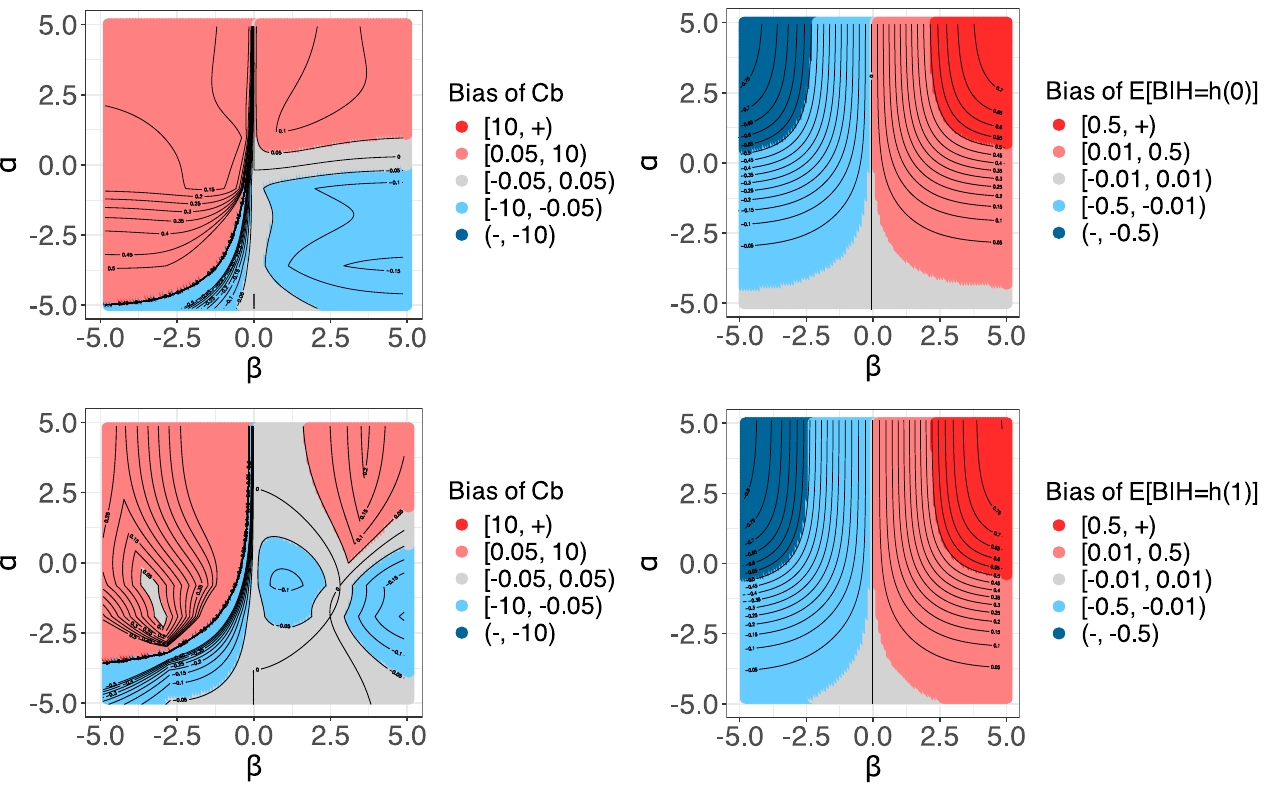}
  \caption[small]{Contour plots of performance metrics. Left: bias of $C_b$ for $p = (0.3, 0.1, 0.4, 0.2)$ (row 1) and $p = (0.300, 0.008, 0.194, 0.498)$ (row 2). Right: bias of the moderate calibration curve for $p = (0.3, 0.1, 0.4, 0.2)$.}
  \label{fig:3}
\end{figure}

Figure \ref{fig:1} displays the moderate calibration curves of all TBPs with and without bias. 
When we fully controls for confounding, $h_2$ and $h_3$ are moderately calibrated as expected, aligning with the 45-degree line. 
The bias of the moderation calibration curve is positive for all four TBPs, due to $\text{bias}(x_1, x_2) > 0$, for all $x_1$ and $x_2$. 
The moderate calibration curve for $h_4$ shows that a TBP {\em developed} in the target population, without controlling for confounder $Z$ during development and evaluation, can falsely suggest moderate calibration.
These findings underscore that not fully controlling for confounding variables can result in misleading patterns. See Appendix S3 for a sanity check in the absence of confounding.

In the second population, the strength of confounding can be adjusted by parameters $\alpha$ and $\beta$, where $\alpha$ controls how $Z$ influences the outcome, while $\beta$ governs how $Z$ affects treatment assignment. We evaluate the optimal TBP, where $h(x) := \tau_0(x)$. Figure~\ref{fig:3} displays a finite set of $(\beta, \alpha)$ pairs, ensuring that the bias of the moderate calibration curve for $h(x)$ equals $\text{bias}(x)$. Continuous contour curves are overlaid to better illustrate the bias structure, with $\alpha$ and $\beta$ varying over the interval $(-5, 5)$. The two plots on the left show the bias of $C_b$ under two joint distributions of $(X,Z)$.
These patterns are not approximately symmetric with respect to the strength of confounding, and they change noticeably when we alter the joint distribution of $(X,Z)$. These two key observations highlight the unpredictable pattern of the relationship between bias and strength of confounding. However, patterns of the moderate calibration curve are more interpretable compared to that of $C_b$, as shown in the two plots on the right. Appendix S4 presents evaluation results for an additional synthetic population with continuous covariates and a continuous outcome, using the same causal structure as Figure~\ref{fig:0}, and details the calculation process.

\section*{Discussion} 
%

In clinical settings, TBPs may offer valuable guidance for physicians and patients in making informed treatment decisions. 
Before being transported to a new population, these TBPs need to be evaluated in that population.
Observational data, where treatment assignment is not random, might be the only opportunity for such evaluation. 
Consequently, addressing confounding bias is crucial when assessing TBPs based on observational data from the target population.

This study demonstrated conceptually how to evaluate pre-specified TBPs, developed from one population, using observational data from a different target population.
We delved into two specific measures, the $C_b$ index and the moderate calibration curve, which are estimands defined in terms of the latent benefit variable $B$.
We showed the identification of $C_b$ and $\operatorname{E}[B \mid H]$ by re-expressing the respective estimands in terms of $\mathbb{P}_{obs}$. This framework supports multiple estimation methods.

The primary role of estimation methods is to estimate $\tau(x, z)$ for all possible values of $x$ and $z$ using sample data.
Estimating $\tau(x, z)$ is a classical topic in the casual inference literature.
For instance, it can be estimated through outcome regression, inverse probability weighting \cite{athey2015machine}, or a combination of both, as per augmented inverse probability weighting \cite{bang2005doubly}.
Once a scheme to estimate $\tau(x, z)$ is chosen, a second layer of estimation is required to target the estimands describing the predictive performance of TBP, such as $C_b$ and the calibration curve.
For instance, one possible estimator of $C_b$ uses sample averages to estimate expectations in the estimand, and then estimates the CDF of $H$ through the empirical distribution of $H$.
For the calibration curve, the sample average within groups sharing the same $H$ can be used to estimate the conditional expectation of the estimated $\tau(X, Z)$ given $H = h$ when $H$ is discrete. For continuous $H$, one approach is to discretize $H$ into equally sized bins and use the sample average within each bin to estimate the conditional expectation\cite{xu2022calibration}.
Moderate calibration can be assessed non-parametrically without grouping or smoothing \cite{sadatsafavi2024non}, though this cumulative error approach has not yet been applied to TBPs. Addressing the challenges of estimating different performance measures remains worthwhile.

The absence of full confounding control leads to bias in identifying $C_b$ and $\operatorname{E}[B \mid H = h]$ for all $h$. 
This study explored the biases of performance metrics and their behavior under positive $\text{bias}(x)$ functions.
A positive $\text{bias}(x)$ function always results in a positive bias in the moderate calibration curve, $\operatorname{E}[B]$, and $\operatorname{E}[BF_H(H)]$; however, it may lead to underestimation of $C_b$ for some choices of $h(x)$. Overestimation of $C_b$ is also possible when the overestimation of $2\operatorname{E}[BF_H(H)]$ exceeds that of $\operatorname{E}[B]$.
When a $\text{bias}(x)$ function takes both positive and negative values, behaviors of bias in identifying $C_b$ and $\operatorname{E}[B \mid H]$ become more unpredictable, making it difficult to intuit the specific direction of bias. 
We also illustrated how confounder strength influences bias in performance metrics using the second population. This process closely aligns with sensitivity analysis, particularly when the single binary variable $Z$ represents an unmeasured confounder. In this case, the parameters $\alpha$ and $\beta$ serve as sensitivity parameters, allowing us to partially recover the performance metrics from the observed data and the plausible range for the sensitivity parameters. The causal structure we considered is simple and, by design, excludes collider bias. In more complex settings, if no sufficient adjustment set that includes variables in $X$ can be identified from the available data, confounding and collider bias may arise and distort evaluation.

In summary, we presented a conceptual framework for evaluating the predictive performance of TBPs using observational data. 
We demonstrated that in the absence of complete confounding control, the unpredictable nature of bias can significantly impact the reliability of TBP assessments.

\pagebreak
\section*{Appendix S1: Propositions for $C_b$ Calculation} 
\addcontentsline{toc}{section}{Appendix S1: Propositions for $C_b$ Calculation}

\begin{proposition}
  Given two independent identical distributed copies, denoted as $\{(B_1, H_1), (B_2, H_2)\}$, expected benefit of the `treat greater $H$' strategy is expressed as: 
  $$\operatorname{E}[B_1I(H_1 \geq H_2) + B_2I(H_1 < H_2)] = \operatorname{E}[B\eta(H)],$$
  where $I(\cdot)$ is the indicator function, and $\eta(H) = 2F_H(H) - f_H(H)$.
  Here, both  $F_H(H)$ and $f_H(H)$ are random variables, with $F_H(\cdot)$ denoting the cumulative distribution function (CDF), and $f_H(\cdot)$ the probability mass function (PMF) of $H$ included as a correction term.
\end{proposition}

\begin{proof}
  We first show that $\eta(H) = 2F_H(H)$ for continuous $H$.
  Without loss of generality, we assume that $B$ is continuous.
  \begin{align*}
      &\operatorname{E}[B_1I(H_1 \geq H_2) + B_2I(H_1 < H_2)] \\
      &= 2\operatorname{E}\big[\operatorname{E}[B_1I(H_1 \geq H_2) \mid H_1, H_2]\big]  \quad \text{($H$ is continuous)}\\
      &= 2\int_{B_1}\int_{H_1} F_{H_2}(h_1) b_1  f_{B_1\mid H_1}(b_1 \mid h_1)f_{H_1}(h_1)dh_1 db_1 \quad \text{($f_{B_1 \mid H_1, H_2} = f_{B_1 \mid H_1}$)}\\
      &=2\int_{B_1}\int_{H_1} b_1F_{H_1}(h_1) f_{B_1, H_1}(b_1, h_1)dh_1db_1 \quad \text{($F_{H_1} = F_{H_2}$)}\\
      &= 2\operatorname{E}[BF_H(H)],
  \end{align*}
    where $f_{B_1, H_1, H_2}(b_1, h_1, h_2)$ is the joint probability density function (PDF) of $(B_1, H_1, H_2)$.

  Then, we show that $\eta(H) = 2F_H(H) - f_H(H)$ for discrete $H$.
  Here, we assume that $B$ is discrete without loss of generality.
  \begin{align*}
    &\operatorname{E}[B_1I(H_1 \geq H_2) + B_2I(H_1 < H_2)] \\
    &= 2 \operatorname{E}\left[E[B_1I(H_1 > H_2) \mid H_1, H_2]\right] + \operatorname{E}\left[ E[B_1I(H_1 = H_2) \mid H_1, H_2]\right]\\
    &= 2\sum_{h_1}\sum_{b_1} b_1\operatorname{P}(B_1 = b_1, H_1 = h_1) \left(\sum_{h_2}I(h_2 < h_1) \operatorname{P}(H_2 = h_2)\right) + \\
    &\sum_{h_1}\sum_{b_1} b_1 \operatorname{P}(B_1 = b_1, H_1 = h_1)\left(\sum_{h_2}I(h_2 = h_1)\operatorname{P}(H_2 = h_2)\right)\\
    &= 2\operatorname{E}[BF_H(H)] - \operatorname{E}[Bf_H(H)].
\end{align*}
Therefore, we have shown that $\operatorname{E}[B_1I(H_1 \geq H_2) + B_2I(H_1 < H_2)] = \operatorname{E}[B\eta(H)]$.
\end{proof}

 For variables $B$ and $H$, the relative concentration curve is $R(p) = \frac{\operatorname{E}[BI(H \leq h)]}{\operatorname{E}[B]}$,
  where $p$ represents the $p$-th quantile concerning the value of $H$.
  We assume that $E[B] > 0$.
\begin{proposition}\label{prop:Gini_like_cont} 
  The Gini-like coefficient is twice the area ($A$) between the line of independence ($p$) and $R(p)$, which satisfies
  $$2A =  \frac{E[B\eta(H)] - \operatorname{E}[B]}{\operatorname{E}[B]}.$$
\end{proposition}

\begin{proof} 
  For continuous $H$, we start with twice the area $p$ and $R(p)$ multiplied by $\operatorname{E}[B]$, which is
      \begin{align*}
          2A\operatorname{E}[B] &= 2\int^{1}_{0} \left(p\operatorname{E}[B] - \int^{h_p}_{-\infty} \operatorname{E}[B \mid H = h]f_H(h)dh \right)dp \\
          &= \operatorname{E}[B] - 2\int^{1}_{0} \int^{h_p}_{-\infty} \operatorname{E}[B \mid H = h]f_H(h)dhdp\\
          &= \operatorname{E}[B] - 2\int^{\infty}_{-\infty} \operatorname{E}[B \mid H = h]f_H(h) (1 - F_H(h))dh\\
          &= 2\int^{\infty}_{-\infty} \operatorname{E}[B \mid H = h]f_H(h)F_H(h)dh - \operatorname{E}[B] \\
          &= 2\operatorname{E}[BF_H(H)] - \operatorname{E}[B],
      \end{align*}
      where $h_p$ represents $h$ value at the $p$-th quantile.
      Since $\eta(H) = 2F_H(H)$ for continuous $H$, we obtain $$2A  = \frac{2\operatorname{E}[B\eta(H)] - \operatorname{E}[B]}{\operatorname{E}[B]}.$$

      For discrete $H$, we assume that $H$ has finite $k$ distinct values, patients are ranked by their value of $H$ in ascending order to plot the relative concentration curve:
      $h_{(1)} < h_{(2)} < h_{(3)} < \cdots < h_{(k)}$ with the probability $\operatorname{P}(H = h_{(i)}) = p_i$, where $i = 1, 2, 3, \cdots, k$ and $\sum_{i=1}^k p_i = 1$.
      We have
  \begin{align*}
   \operatorname{E}[BF_H(H)] &= \left(p_1\operatorname{E}[BI(H = h_{(1)})] + (p_1 + p_2)\operatorname{E}[BI(H = h_{(2)})] + \cdots + 1 \cdot \operatorname{E}[I(H = h_{(k)})]\right),\\
   \operatorname{E}[Bf_H(H)] &= \sum^k_{i=1} p_i \operatorname{E}[BI(H = h_{(i)})].
  \end{align*}
  When $B \geq 0$, area $A$ would be bounded between $0$ and $0.5$, which can be calculated as $0.5$ minus the sum of areas of one triangle and $(k-1)$ trapezoids.
  Therefore, we can express the Gini-like coefficient as
  \begin{align*}
      2A &= \frac{1}{\operatorname{E}[B]} \left((1-p_k)\operatorname{E}[B] - \sum^{k-1}_{i=1} (p_i+p_{i+1})\operatorname{E}[BI(H\leq h_{(i)})]\right),
  \end{align*}
  where $\operatorname{E}[BI(H\leq h_{(k)})] = \sum_{i = 1}^k \operatorname{E}[BI(H = h_{(i)})]$.
  Then, we can find that $\operatorname{E}[BF_H(H)]$, $\operatorname{E}[Bf_H(H)]$, $\operatorname{E}[B]$, and $2A$ satisfy the relationship
  \begin{align*}
      \frac{2\operatorname{E}[BF_H(H)] - \operatorname{E}[B]}{\operatorname{E}[B]} - 2A&= \frac{\operatorname{E}[Bf_H(H)]}{\operatorname{E}[B]}.
   \end{align*}
Since $\eta(H) = 2F_H(H) - f_H(H)$ discrete $H$, we obtain $$2A  = \frac{2\operatorname{E}[B\eta(H)] - \operatorname{E}[B]}{\operatorname{E}[B]}.$$

When distribution of $B$ includes negative values or `treat greater $H$' is worse than `treat at random,' this relationship still holds.
However, the area $A$ can take a value greater than $0.5$ or less than $0$, causing $C_b$ to fall outside the interval $(0, 1)$.
This issue also arises with the Gini coefficient and the Lorenz curve, and it has been discussed in the literature.
\end{proof}

\section*{Appendix S2: Synthetic Populations Setup}
\addcontentsline{toc}{section}{Appendix S2: Synthetic Populations Setup}

For simplicity, we assume that $Y^{(0)} \indep Y^{(1)} \mid A, Z, X$, where $X$ is a covariate vector $X = (X_1, X_2)$.
\begin{itemize}
    \item Target population 1:
\begin{align*}
    (Y^{(0)} \mid X_1 = x_1, X_2 = x_2, Z = z) &\sim \text{Bernoulli}(0.6290 + 0.1434x_1 -0.4794x_2 -0.0579z),\\
    (Y^{(1)} \mid X_1 = x_1, X_2 = x_2, Z = z) &\sim \text{Bernoulli}(0.3348 + 0.3037x_1 -0.3338x_2 +  0.3138z),\\
    (A \mid Z = z)  &\sim \text{Bernoulli}(0.1204 + 0.7621z),\\
    (X_1, X_2, Z) &\sim \text{Multivariate}\big(p\big),
\end{align*}
where $p = (p_{111}, p_{110}, p_{101}, p_{100}, p_{011}, p_{010}, p_{001}, p_{000})$ and we set the vector of $8$ parameters as $(0.1807, 0.1005, 0.0354, 0.1482, 0.1745, 0.0872, 0.1208, 0.1528)$.
     \item Target population 2:
     \begin{align*}
    (Y^{(0)} \mid X = x, Z = z) &\sim \text{Bernoulli}(\text{expit}(x + \alpha z)),\\
    (Y^{(1)} \mid X = x, Z = z) &\sim \text{Bernoulli}(\text{expit}(0.1 + x + \alpha z)),\\
    (A \mid z)  &\sim \text{Bernoulli}(\text{expit}(0.35 + \beta z)),\\
    (X, Z) &\sim \text{Multivariate}\big(p\big),
\end{align*}
where $p = (p_{11}, p_{10}, p_{01}, p_{00})$.
\end{itemize}

\textit{Note that all reported values are rounded to four decimal places; totals may not sum to $1.0000$ due to rounding. All calculations used full-precision values.}
\section*{Appendix S3: Supplementary Information for Population 1}
\addcontentsline{toc}{section}{Appendix S3: Supplementary Information for Population 1}

In the first population example, we defined a distribution $\mathbb{P}$ and a linear function $\tau(x_1, x_2,z)$ using a total of 18 parameters. 
To begin with, we design moderately and strongly calibrated TBPs. 
To design a moderately calibrated $h_2(x_1,x_2)$, we specific a vector of coefficients:
\begin{align*}
    c_0 &= \left[ (\alpha_{10} - \alpha_{00}) + (\alpha_{13} - \alpha_{03})\frac{p_{001}}{p_{001} + p_{000}}\right], \\ 
    c_1 &= c_2 = \Bigg[(\alpha_{11} - \alpha_{01})\frac{p_{101} + p_{100}}{p_{101} + p_{100} + p_{011} + p_{010}} + (\alpha_{12} - \alpha_{02})\frac{p_{011} + p_{010}}{p_{101} + p_{100} + p_{011} + p_{010}} +\\
    &(\alpha_{13} - \alpha_{03}) \left( \frac{p_{101} + p_{011}}{p_{101} + p_{100} + p_{011} + p_{010}} - \frac{p_{001}}{p_{001} + p_{000}} \right)\Bigg],\\
    c_3 &=  \Bigg[(\alpha_{11} - \alpha_{01})\left(1 - 2 \frac{p_{101} + p_{100}}{p_{101} + p_{100} + p_{011} + p_{010}} \right)+ \\
    &(\alpha_{12} - \alpha_{02})\left(1 - 2\frac{p_{011} + p_{010}}{p_{101} + p_{100} + p_{011} + p_{010}}\right) +\\
    &(\alpha_{13} - \alpha_{03}) \left( \frac{p_{111}}{p_{111} + p_{110}} - 2\frac{p_{101} + p_{011}}{p_{101} + p_{100} + p_{011} + p_{010}} + \frac{p_{001}}{p_{001} + p_{000}} \right)\Bigg].
\end{align*}
To design a strongly calibrated $h_3(x_1,x_2)$, we set 
\begin{align*}
    c_0 &= \left[ (\alpha_{10} - \alpha_{00}) + (\alpha_{13} - \alpha_{03})\frac{p_{001}}{p_{001} + p_{000}}\right], \\ 
    c_1 &= \left[(\alpha_{11} - \alpha_{01}) + (\alpha_{13} - \alpha_{03}) \left( \frac{p_{101}}{p_{101} + p_{100}} - \frac{p_{001}}{p_{001} + p_{000}} \right)\right],\\
    c_2 &=  \left[(\alpha_{12} - \alpha_{02}) + (\alpha_{13} - \alpha_{03}) \left( \frac{p_{011}}{p_{011} + p_{010}} - \frac{p_{001}}{p_{001} + p_{000}} \right)\right],\\
    c_3 &=  \left[(\alpha_{13} - \alpha_{03}) \left( \frac{p_{111}}{p_{111} + p_{110}} - \frac{p_{101}}{p_{101} + p_{100}} - \frac{p_{011}}{p_{011} + p_{010}} + \frac{p_{001}}{p_{001} + p_{000}} \right)\right].
\end{align*}

\begin{figure}[h]
  \centering
  \includegraphics[scale=0.5]{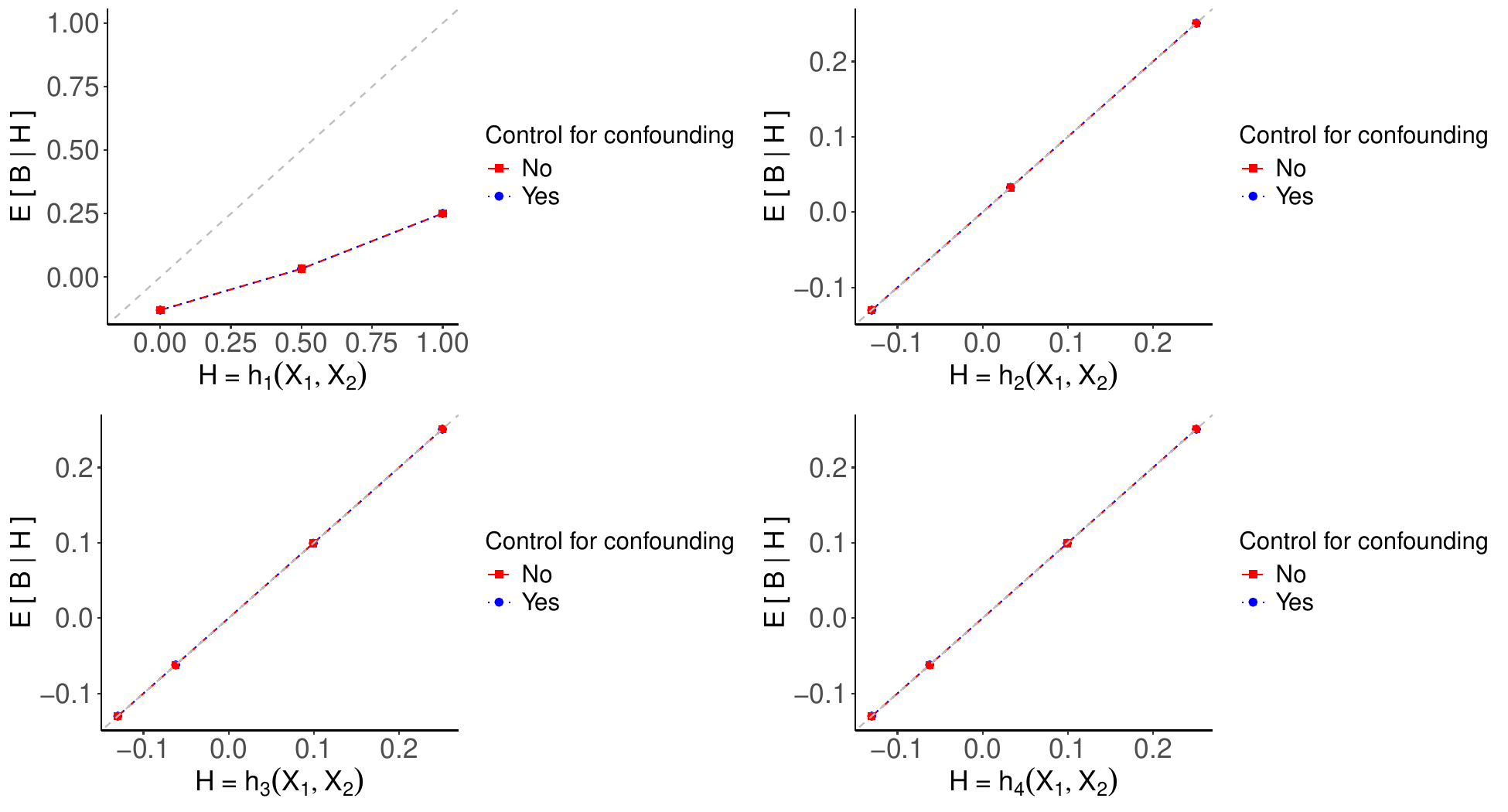}
  \caption[Evaluation Results for Population 1]{The moderate calibration plots for the four TBPs when $\beta_1 = 0$.
  The blue dotted curves refer to $\operatorname{E}[B \mid H]$, and the red dashed curves refer to $\tilde{\operatorname{E}}[B \mid H]$.}
  \label{fig:4}
\end{figure}

Finally, we examine the discrimination and calibration performance of $h_1$, $h_2$, $h_3$, and $h_4$ in the absence of confounding bias.
When $\beta_1 = 0$, we have $\text{bias}(X) = 0$, resulting in zero bias in estimating both $C_b$ and $\operatorname{E}[B \mid H = h]$, for all $h$.
The results are presented in Table~\ref{tab:cb0} and Figure~\ref{fig:4}.
In particular, the red and blue calibration curves for $h_2(x_1, x_2)$, $h_3(x_1, x_2)$, and $h_4(x_1, x_2)$ align with the 45-degree diagonal line.

\begin{table}[ht!]
  \centering
  \caption{Values of $C_b$ with and without full confounding control}
  \label{tab:cb0}
  \begin{tabular}{l c c c c}
    \toprule
    Metric & $ h_1(x_1,x_2) $ & $ h_2(x_1,x_2) $ & $ h_3(x_1,x_2) $ & $ h_4(x_1,x_2) $\\
    \midrule
        $C_b$        & 0.6069 & 0.6069 & 0.6298 & 0.6298  \\
    $\tilde{C}_b$     &0.6069 & 0.6069 & 0.6298 & 0.6298  \\
    \bottomrule
  \end{tabular}

  {\footnotesize Note: $C_b$ represents correct evaluation values and $\tilde{C}_b$ represents values with bias.}
\end{table}

\section*{Appendix S4: Closed-form Measure Calculations for an Additional Population}
\addcontentsline{toc}{section}{Appendix S4: Closed-form Calculations for an Additional Population}

In this synthetic population, obesity, symptom severity, and socioeconomic status are independent and continuous, each following a uniform distribution on the interval $[0,1]$. The exposure is binary, and the probability of receiving bronchodilator therapy is directly influenced by the values of socioeconomic status. Outcome $\text{FEV}_1$ is continuous, which has non-linear relationships with the exposure and covariates. Let $X = (X_1, X_2)$ denote the covariate vector with realization $x = (x_1,x_2)$.
Assume $Y^{(0)} \indep Y^{(1)} \mid A, Z, X$, and 
    \begin{align*}
    (Y^{(a)} \mid X=x, Z = z)  &\sim \text{N}(\tau_0(x)\left(a - 0.5\right) + b(x,z), 0.01),\\
     (A \mid Z = z) &\sim \text{Bernoulli}(z),
     \end{align*} 
     where $X, Z \overset{iid}{\sim} \text{Unif}(0,1)$, $\tau_0(x) = \max \left(x_1, x_2\right)$ and $b(x,z) = \max \left(z, x_2\right) + 0.1x_1$.
The conditional expectation of index improvement conditional on $X = x$, and $Z = z$ are defined by two functions  for the control and treatment groups. For the control group, it equals minus half of $\tau_0(x)$ plus a base response function $b(x, z)$. For the treatment group, It equals half of $\tau_0(x)$ plus the same base response function, where $\tau_0(x)$, is defined as the maximum value between obesity and symptom severity. The base response function, $b(x, z)$, is a nonlinear function.

A similar setup has been used by Foster and Syrgkanis \cite{foster2023orthogonal}. From these population specifications, we can derive that $\tau(x, z) = \tau_0(x)$, for all $x$ and $z$, which shows that socioeconomic status contributes to explaining both the outcome and therapy assignment but is independent of treatment benefit conditional on $X = x$. 

We first compute $\operatorname{E}[B \mid H = h]$.
Given that $\operatorname{E}[B \mid H = h] = \operatorname{E}[\tau_0(X) \mid H = h]$, we need the joint PDF of $(\tau_0(X), H)$, where $\tau_0(X) = \max(X)$ and $H = X_1 + X_2$.
Note that $H$ follows a triangular distribution with lower limit $a = 0$, upper limit $b = 2$, and mode $c = 1$. 
Since $\tau_0$ is not a one-to-one function, we examine the pre-image of a point $(\tau_0(x), h)$.
When $h \leq 2\tau_0(x)$ and $\tau_0(x) \leq h$, the pre-image of any point $(\tau_0(x), h)$ consists of two values, which are $(x_1 = \tau_0(x), x_2 = h - \tau_0(x))$ and $(x_1 = h - \tau_0(x), x_2 = \tau_0(x))$.
These two points correspond to two scenarios: $x_1 \geq x_2$ and $x_1 < x_2$. 
In each scenario, max function is a one-to-one mapping.

Therefore, the joint PDF of $(\tau_0(X), H)$ is
$$f_{\tau_0(X), H}(\tau_0(x),h) = \begin{cases} 
    2, &\tau_0(x) \leq h , h \leq 2\tau_0(x), 0 \leq \tau_0(x) \leq 1, \\
    0, & \text{otherwise}.
 \end{cases}$$
With the expression of $f_{\tau_0(X), H}(\tau_0(x),h)$, we can calculate two marginal PDFs to check that the random variable $\tau_0(X)$ follows the $\text{Beta}(2,1)$ distribution and the random variable $H$ follows the $\text{triangular}(a = 0, b = 2, c = 1)$ distribution.
Thus, the CDF of $H$ is 
\begin{align*}
  F_{H}(h) = \begin{cases} 
      0, &  h < 0,\\
      h^2/2, &  0 \leq h < 1, \\
      1 - (2-h)^2/2, &  1 \leq h < 2, \\
      1, & 2 \leq h.
    \end{cases}
\end{align*}
With $f_{\tau_0(X), H}(\tau_0(x),h)$, we calculate the conditional PDF of $(\tau_0(X) \mid H)$:
\begin{align*}
    f_{\tau_0(X) \mid H}(\tau_0(x) \mid h)  &=\begin{cases} 
        2/h, & 0 < h \leq 1, \tau_0(x) \leq h, h \leq 2\tau_0(x), \\
        2/(2 - h), & 1 < h < 2, \tau_0(x) \leq h, h \leq 2\tau_0(x),  \tau_0(x) \leq 1, \\
        0, & \text{otherwise}.
     \end{cases}
\end{align*}
We obtain the moderate calibration curve:
\begin{align*}
  \operatorname{E}[\tau_0(X) \mid H = h] & = \int_{\tau_0(X)} \tau_0(x) f_{\tau_0(X) \mid H}(\tau_0(x) \mid h) d\tau_0(x)\\
    &= \begin{cases} 
      \frac{3h}{4}, & 0 < h \leq 1, \\
      \frac{1-h^2/4}{2 - h}, & 1 < h < 2.
   \end{cases}
\end{align*}

Then, we compute $C_b$ for $H$ via the expression $1 - \operatorname{E}[\tau_0(X)]/ 2\operatorname{E}[\tau_0(X)F_{H}(H)]$, where $\operatorname{E}[\tau_0(X)] = 2/3$ as $\tau_0(X)$ follows the $\text{Beta}(2,1)$ distribution.
With the expression of $f_{\tau_0(X), H}(\tau_0(x),h)$, we have 
\begin{align*}
  \operatorname{E}[\tau_0(X)F_{H}(H)] &= \int_{h}\int_{\tau_0(x)} \tau_0(x) F_{H}(h)f_{\tau_0(X), H}(\tau_0(x),h)d\tau_0(x) dh\\
    &= \int^1_{0}\int^{h}_{\frac{h}{2}} 2\tau_0(x) \frac{h^2}{2}d\tau_0(x) dh + \int^2_{1}\int^{1}_{\frac{h}{2}} 2\tau_0(x) \left(1 - \frac{(2-h)^2}{2}\right)d\tau_0(x) dh\\
    &= 2\left(\frac{3}{80} + \frac{19}{120}\right) = E[BF_H(H)].
\end{align*}
Therefore, we have
$$C_{b,h} = 1 - \frac{\operatorname{E}[\tau_0(X)]}{2E[BF_H(H)]} = 1 - \frac{2/3}{4\left(\frac{3}{80} + \frac{19}{120}\right)} = 0.1489.$$
Furthermore, we can also calculate the $C_b$ for $\tau_0(X)$ following the same processes, which is $1 - \frac{2/3}{4/5} = 0.1667$.
In this setup, $C_b = 0.1489$ indicates that assigning treatment to a patient who has a larger $H$ value is slightly better than random treatment assignment.

\begin{figure}[h]
  \centering
  \includegraphics[scale=0.5]{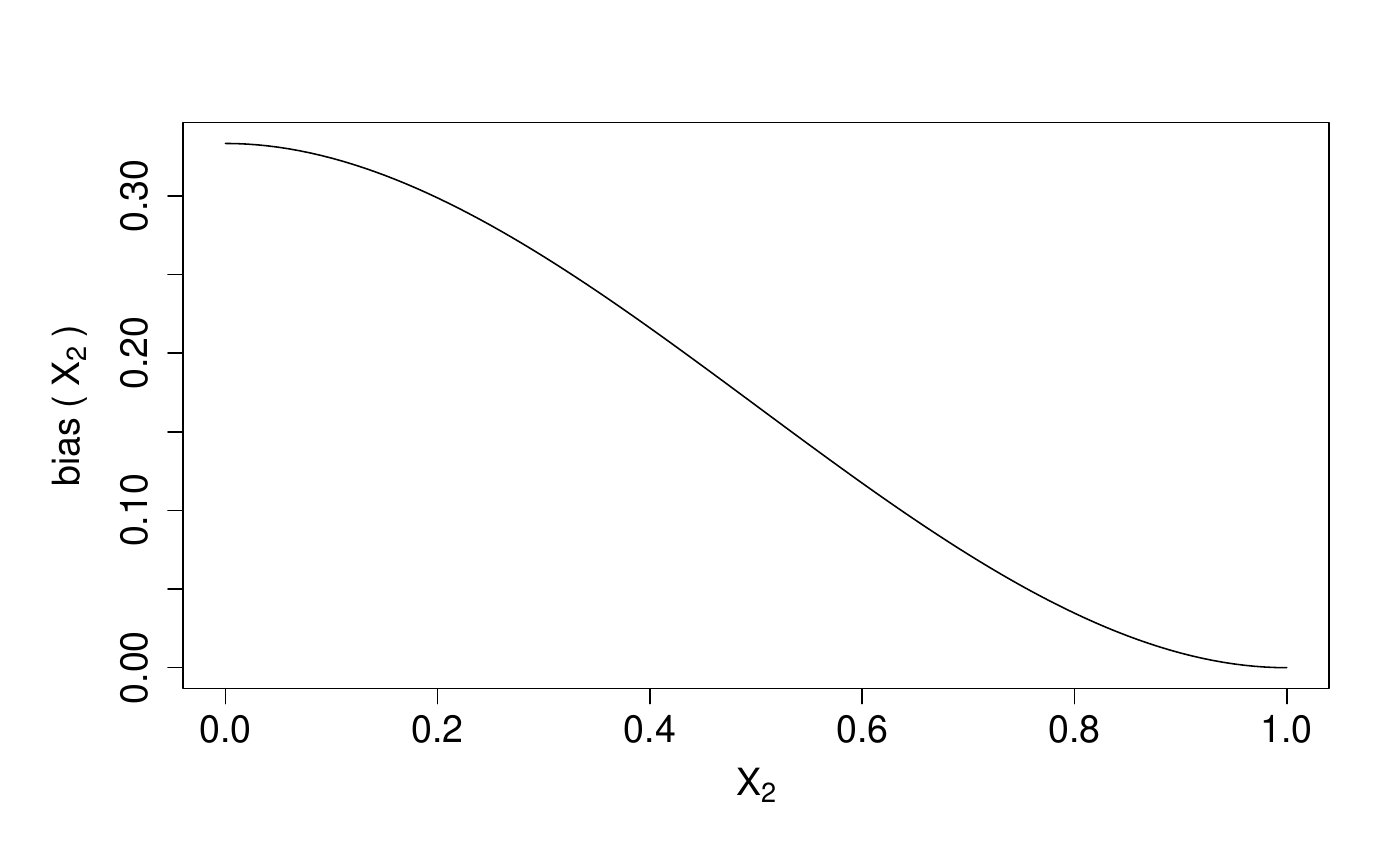}
  \caption[The Confounding Bias Function $\text{bias}(X)$]{The confounding bias function, $\text{bias}(X)$, which is only a function of $X_2$.}
  \label{fig:5}
\end{figure}

Finally, we compute $\tilde{\operatorname{E}}[B \mid H = h]$ and $\tilde{C}_b$. Recall that the confounding bias is defined as
\begin{align*}
    \text{bias}(x) = \left(\mu_{1}(x) - \mu_{0}(x)\right) - \tau_0(x).
\end{align*}
Given $\mu_{a}(x,z)$ for $a \in \{0,1\}$, we obtain $\mu_{a}(x)$:
\begin{align*}
  \mu_{1}(x) &= \operatorname{E}[Y \mid A = 1, X = x] \\
  &= \int_Z \operatorname{E}[Y \mid A = 1, X = x, Z = z] f_{Z \mid X, A}(z \mid x, 1)dz\\
    &= \int_Z 2z\left(\frac{1}{2}\max(x_1,x_2) + \max(x_2,z) + \frac{1}{10}x_1\right)dz\\
    &= \frac{1}{2}\max(x_1,x_2) + \frac{1}{3}x_2^3 + \frac{2}{3} + \frac{1}{10}x_1\\
 \mu_{0}(x) &= \operatorname{E}[Y \mid A = 0, X = x]\\
 &= \int_Z \operatorname{E}[Y \mid A = 0, X = x, Z = z] f_{Z \mid X, A}(z \mid x, 0)dz\\
    &= \int_Z 2(1 - z)\left(-\frac{1}{2}\max(x_1,x_2) + \max(x_2,z) + \frac{1}{10}x_1\right)dz\\
    &= -\frac{1}{2}\max(x_1,x_2) +  \frac{1}{3} + x_2^2 -  \frac{1}{3}x_2^3 + \frac{1}{10}x_1.
\end{align*}
Denote $D(x) = \mu_{1}(x) - \mu_{0}(x)$, and we have
\begin{align*}
    D(x) &=  \tau_0(x_1,x_2) + \frac{1}{3} - x_2^2 + \frac{2}{3}x_2^3\\
    &= \tau_0(x) + \text{bias}(x),
\end{align*}
where $\text{bias}(x) = \frac{2}{3}x_2^3 - x_2^2 + \frac{1}{3}$. This confounding bias is illustrated in Figure~\ref{fig:5}. 
To compute $\tilde{\operatorname{E}}[B \mid H = h]$, we express 
\begin{align*}
  \operatorname{E}[D(X) \mid H = h] &= \operatorname{E}[\tau_0(X) \mid H = h] + \operatorname{E}[\text{bias}(X) \mid H = h],
\end{align*}
where $\operatorname{E}[\tau_0(X) \mid H = h]$ has been calculated.
To calculate $\operatorname{E}[\text{bias}(X) \mid H = h]$, we need to figure out the joint distribution of $(X_2, H)$ and then the conditional distribution of $X_2$ given $H = h$.
Solving the linear system, we get
$$f_{X_2, H}(x_2, h) = f_{X_2, X_1 + X_2}(x_2, x_1 + x_2) = f_{X_2, X_1}(x_2, x_1)|J| = 1,$$
where $0 \leq x_2 \leq 1$, $0 \leq h \leq 2$, $h - 1\leq x_2$, and $x_2 \leq h$.
The conditional PDF should be
\begin{align*}
    f_{X_2 \mid H}(x_2 \mid h) &=\begin{cases} 
        \frac{1}{h}, & 0 \leq x_2 \leq 1, 0 < h \leq 1, h - 1\leq x_2, x_2 \leq h, \\
        \frac{1}{2 - h}, & 0 \leq x_2 \leq 1, 1 \leq h < 2, h - 1\leq x_2, x_2 \leq h, \\
        0, & \text{otherwise}.
     \end{cases}
\end{align*}
Therefore, we have
\begin{align*}
  \operatorname{E}[\text{bias}(X_2) \mid H = h] 
    &=\begin{cases}
        \frac{1}{6}\left(h^3 - 2h^2 + 2\right), & 0 < h \leq 1,\\
        \frac{1}{6}\left(h^3-4h^2+4h\right), & 1 < h < 2.
    \end{cases}\\
        \tilde{\operatorname{E}}[B \mid H = h]
    &=\begin{cases}
      \frac{3h}{4} + \frac{h^3 - 2h^2 + 2}{6}, & 0 < h \leq 1,\\
      \frac{1-h^2/4}{2 - h} + \frac{h^3-4h^2+4h}{6}, & 1 < h < 2.
    \end{cases}
\end{align*}

Similarly, we calculate the $\tilde{C}_b$ for $H$ by computing 
\begin{align*}
\operatorname{E}[D(X)] &= \operatorname{E}[\tau_0(X)] + \operatorname{E}[\text{bias}(X)]\\
    &= \frac{2}{3} + \int_{x_2} \left(\frac{2}{3}x_2^3 - x_2^2 + \frac{1}{3}\right)f_{X_2} (x_2)dx_2\\
    &=\frac{2}{3} + \frac{1}{6} = \frac{5}{6}.\\
  \operatorname{E}[D(X)F_{H}(H)] &= \operatorname{E}[\tau_0(X)F_{H}(H)] + \operatorname{E}[\text{bias}(X)F_{H}(H)]\\
  &= 2\left(\frac{3}{80} + \frac{19}{120} \right) + \int_{x_2}\int_{h} \left(\frac{2}{3}x_2^3 - x_2^2 + \frac{1}{3}\right)  F_{H}(h)f_{X_2, H}(x_2,h)dx_2 dh\\
    &= 2\left(\frac{3}{80} + \frac{19}{120} \right) + \left(\frac{13}{504} + \frac{43}{1260}\right).
\end{align*}
Therefore,
\begin{align*}
    \tilde{C}_{b,h} = 1 - \frac{\operatorname{E}[D(X)]}{2\operatorname{E}[D(X)F_{H}(H)]} = 1 - \frac{5/6}{2\left(2\left(\frac{3}{80} + \frac{19}{120} \right) + \left(\frac{13}{504} + \frac{43}{1260}\right)\right)} = 0.0773,
\end{align*}
which is lower than $C_b$.

 \begin{figure}[h!]
    \centering
    \includegraphics[scale=0.5]{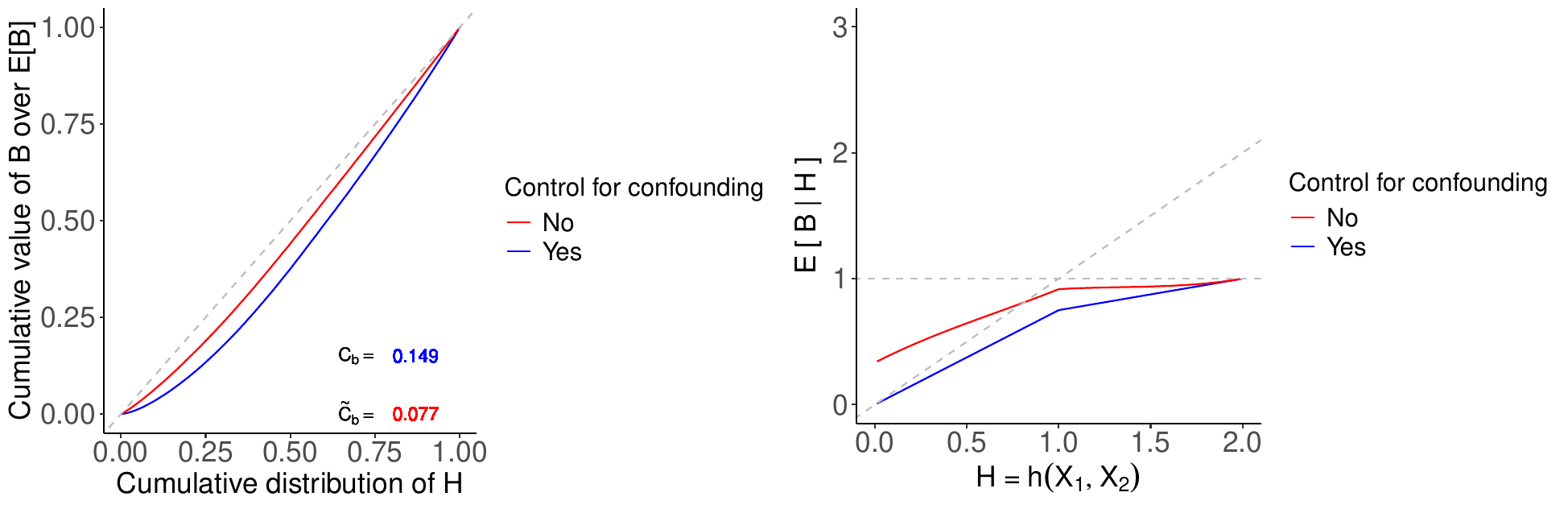}
    \caption[Evaluation Results for an Additional Population]{The relative concentration curves, the $C_b$ indices (left) and moderate calibration curves (right).}
    \label{fig:2}
  \end{figure}
We have shown that confounding bias causes deviations from the actual $C_b$ and the moderate calibration curve, which are also depicted in Figure~\ref{fig:2}.
The bias reduces the area between the independence line and the relative concentration curve by roughly half. It causes an overestimation of $\operatorname{E}[B]$ and an overestimation of $2\operatorname{E}[BF_H(H)]$, but an underestimate $C_b$. In the calibration plot, the red curve deviates from the blue curve as the value of $H$ approaches zero. However, this deviation diminishes to zero as $H$ approaches two.

\pagebreak
\bibliographystyle{ama} 
\bibliography{biblio.bib}

\end{document}